\documentclass[letterpaper]{jpconf}
\usepackage{graphicx}
\begin{document}
\title{The Milky Way's stellar halo - lumpy or triaxial?}

\author{Heidi Jo Newberg$^1$ and Brian Yanny$^2$}

\address{$^1$ Department of Physics, Applied Physics and Astronomy, Rensselaer Polytechnic Institute, Troy, NY 12180} 
\address{$^2$ Fermi National Accelerator Laboratory, P.O. Box 500, Batavia, IL 60510}

\ead{heidi@rpi.edu, yanny@fnal.gov}

\begin{abstract}
We present minimum chi-squared fits of power law and Hernquist density 
profiles to F-turnoff stars in eight 2.5$^\circ$ wide stripes of SDSS data: 
five in the North Galactic Cap and three in the South Galactic cap.  Portions 
of the stellar Galactic halo that are known to contain large streams of 
tidal debris or other lumpy structure, or that may include significant contamination
from the thick disk, are avoided.  The data strongly favor a model that is
not symmetric about the Galaxy's axis of rotation.  If included as a free
parameter, the best fit to the center of the spheroid is surprisingly $\approx 3$kpc 
from the Galactic center in the direction of the Sun's motion.  The model fits 
favor a low value of the density of halo stars at the solar position.  The alternative
to a non-axisymmetric stellar distribution is that our fits are contaminated
by previously unidentified lumpy substructure.
\end{abstract}

\section{Introduction}

The stars in the Milky Way are divided into: an old, metal-rich bulge population near the Galactic
center; a thin disk population in the plane of the Milky Way that includes a range
of
stellar ages and metallicities but is unique in including a young, metal-rich
population; a thick disk 
population of intermediate age and metallicity (fraction of elements heavier
than helium); and an  old, metal-poor spheroid (also called the 
stellar halo) population in which the disks are embedded.  
Each of the populations has a distinct kinematic signature.
The spheroid is generally described as a somewhat spherical population with a
density that falls off as $r^{-\alpha}$, $\alpha=3.5$, where $r$ is the 
distance from the center of the Galaxy.
A squashing parameter $0.5 < q < 1.0$, in the direction perpendicular to the Galactic plane is also fit
to the density distribution of the spheroid by redefining the distance from the Galactic center
as: $r = \sqrt{X^2 + Y^2 + (Z/q)^2}.$

The Sloan Digital Sky Survey (SDSS) scans the sky in $2^\circ.5$ wide stripes 
that follow great circles on the sky.  The photometric images are taken in
five optical filters: $ugriz$.  Data that has been corrected for interstellar
reddening using IRAS dust maps \cite{sfd} is designated by a subscripted
zero.  Three public data releases 
\cite{dr1,dr2,dr3} are available though an online catalog archive server.  
Though the survey geometry and technical design \cite{fukugita, gunn, 
hogg, stoughton, pier, smith, york} optimized extragalactic research, 
studies of the Galactic stars in the photometric catalog have significantly
advanced our knowledge of the stars in the halo of our own Galaxy.

In particular, it has been shown \cite{astrom} that the distribution of faint
F-turnoff stars ($19 < g < 21.5$) that we generally associate with the
spheroid population \cite{besancon, newberg2002} are not distributed 
symmetrically about the Galaxy center.  Even if regions surrounding
known or suspected tidal streams and debris are eliminated from 
consideration, the stellar distribution is not symmetric about a plane
(perpendicular to the Galactic plane) through Galactic longitudes $l=0^\circ, l=180^\circ$.  
Since there are no values of the axisymmetric power law model
parameters ($\alpha,q$) which describe an asymmetric distribution, the
standard model must be modified to include asymmetric stellar populations,
large scale lumpiness, or an asymmetric smooth spheroid component.

\begin{figure}
\begin{center}
\includegraphics[width=4.5in]{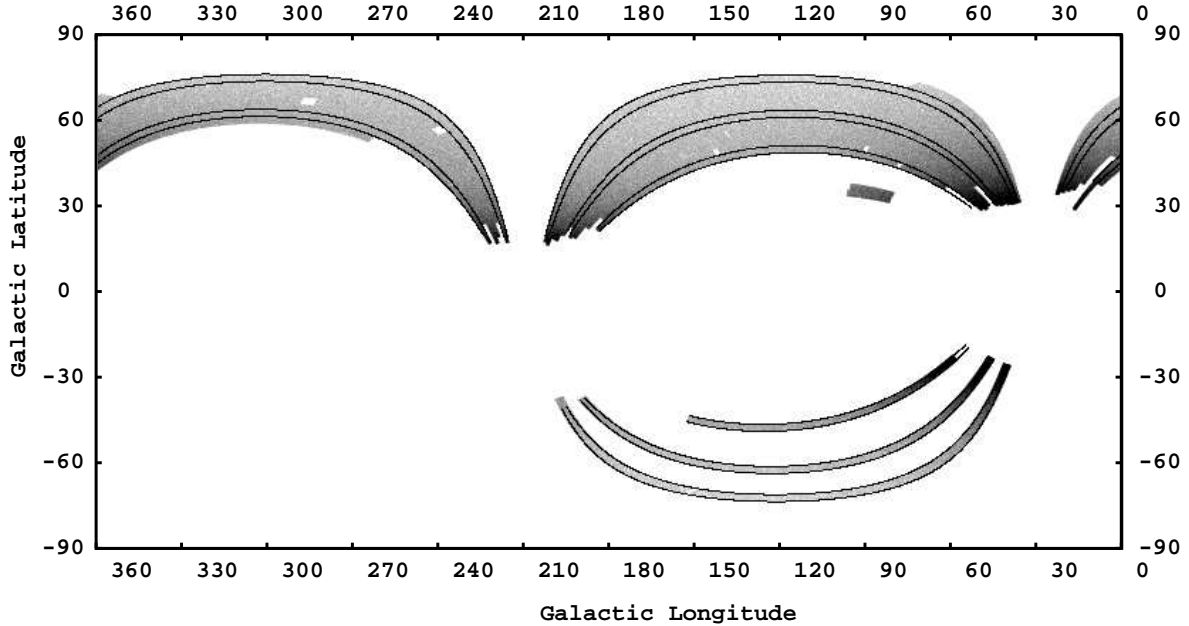}
\end{center}
\caption{\label{stripes} Location of the DR3 data used to fit the
spheroid in Galactic coordinates.  The grey scale shows spheroid stars
selected from DR3 in the manner described in \cite{astrom}.  The black
lines show the outlines of eight stripes used in this
paper to fit the spheroid models.  The SDSS stripes numbered are 
sequentially in a clockwise direction around the coordinate 
$(l,b) = (209^\circ, -7^\circ)$.  This figure shows stripes 10 (bottom) 
and 15 (top) 
in the range $0^\circ < l < 29^\circ, 209^\circ < l < 360^\circ$; and stripes 
24, 32, 37, 76, 82, 86, top to bottom, in the range $29^\circ < l < 209^\circ$.}
\end{figure}

In this paper, we make the assumption that there exists a smooth component
to the Galactic halo, and present several candidate models.  We then
use a minimum $\chi^2$ technique to find the best fit parameters for
each model, fitting to eight $2^\circ.5$ wide stripes of data in the SDSS
DR3.  These stripes are chosen to sample the widest range of positions
available in the Galaxy from this dataset.  Figure~\ref{stripes} shows the 
positions of these data stripes in Galactic coordinates.

\section{Spheroid Models and Parameters}

The simplest extension to the oblate power law spheroid model that allows
for asymmetry about the Galactic center is a triaxial power law spheroid.
In addition to flattening the spheroid by a factor of $q$ in the $Z$
direction, we flatten by a factor of p in the $Y$ direction. 
This density function with ellipsoidal contours can in general be shifted 
in three coordinates and rotated by three angles with respect to the 
Galactic coordinates $(X, Y, Z)$, where $X$ goes from the solar position
through the center of the Galaxy, $Y$ is in the direction of the local
standard of rest (roughly the direction the Sun is going), and we have
a right-handed coordinate system centered on the center of the Milky Way.
The principle axes of the density function are:
\[
\begin{array}{l}
X' = x\cos\theta \cos\phi + y\sin\theta \cos\phi + z\sin\phi\\
Y' = x(-\sin\theta \cos\xi-\cos\theta \sin\phi \sin\xi)+y(\cos\theta \cos\xi-\sin\theta \sin\phi \sin\xi)+z\cos\phi \sin\xi\\
Z' = x(\sin\theta \sin\xi -\cos\theta \sin\phi \cos\xi)+y(-\cos\theta \sin\xi-\sin\theta \sin\phi \cos\xi)+z\cos\phi \cos\xi,\\
\end{array} \]
where $x=X-dX, y=Y-dY,$ and $z=Z-dZ.$
Initially, we expected that $dX=dY=dZ=\phi=\xi=0$, but the addition
of these parameters to the models significantly improved the fits.

We also found that power laws tended to predict too high a star density
near the center of the Galaxy, compared to our data.  We achieved a
better fit to the data with a Hernquist profile:
\[\rho = \frac{\rho_0}{r(r+R_{core})^3}, 
r = \sqrt{X'^2 + (Y'/p)^2 + (Z'/q)^2}.\]
Although we did explore
the possibility of fitting a function of the form:
\[\rho = \frac{\rho_0}{r^\alpha(r+R_{core})^\delta},\]
which would include both the power law model ($\delta=0$) and the Hernquist
model ($\alpha=1, \delta=3$), this fitting was abandoned because we
are not very sensitive to the combination of $\alpha$ and $\delta$.  In
general, models with $\alpha + \delta \approx 4$ worked well.

\section{F Turnoff Star data}

The data includes all stellar objects in the 8 SDSS stripes with:
$(r-i)_0 < 0.8$ (to get rid of 
M stars at faint magnitudes), $0.4 < (u-g)_0 < 3.8$ (to eliminate QSOs), 
$0.0 < (g-r)_0 < 0.3$ (to select stars with turnoff colors), and $g_0< 24.$ 
The designation as ``star" in the database removes issues of duplication of
objects due to overlaps and deblending.  Objects are further rejected if
flags indicate that they are saturated, too near the edge of an image
to give good photometry, or classified as a ``bright" object.  This resulted
in selection of (234,306; 190,221; 159,632; 132,637; 123,359; 115,416; 165,620;
150,562) stars from stripes (10, 15, 24, 32, 37, 76, 82, 86), respectively.
These stripes are shown in figure \ref{stripes}.

In order to compare the stellar density to the model density for a given
set of parameters, the stellar data was sorted into bins that were 0.1 
mag. wide in $g_0$ and $1^\circ$ in angle along the length of the 
stripe.  For a given model, the expected number of stars in each bin
was calculated as:
\[N = \rho(r) dV = \rho(r) R^3 \frac{\ln{10} \Omega dm}{5},\]
where $R$ is the distance from the Sun, $dm$ is the magnitude bin size,
$\Omega$ is the solid angle of each bin (2.5 sq. deg.), and $\rho$ is the 
density of our sample of F turnoff stars.

Since our models included only stars in a smooth spheroid population, we
selected data only from parts of sky and magnitude ranges we thought were
most likely to contain exclusively stars from this population.  Figures
\ref{stripe37}, \ref{stripe10}, \ref{stripe15}, \ref{stripe24}, \ref{stripe32}, 
\ref{stripe76}, \ref{stripe82}, and \ref{stripe86} show the outlines of the
portions of the data that seemed the most likely to contain purely spheroid
stars.  Bright magnitudes were avoided, particularly near the Galactic
plane, since they could be contaminated with thick disk stars (as determined
by Besancon galaxy models \cite{besancon}).  Stars
near any previously identified spheroid overdensity were also avoided.

The models were then modified to reflect the uncertainty in the absolute
magnitudes of the turnoff stars, which leads to an uncertainty in the
distance to each star.  We analyzed the stars in the three globular
clusters Pal 5 (stripe 10), M15 (stripe 76), and M2 (stripe 82) to
determine the turnoff star counts as a function of apparent
magnitude.  A histogram of star counts vs. $g_0$ magnitude was created
by subtracting the star counts in a neighboring piece of sky from the
star counts in the direction of the globular cluster.  A function of
the form: 
\[
f(g_0) = A e^{-(g_0-g_{peak})/(2 \sigma^2)} \left\{
\begin{array}{ll}
\sigma = 0.55, & g_0<g_{peak}\\
\sigma = 0.75, & g_0>g_{peak}
\end{array} \right. \]
was a reasonable fit to the data.  This is an approximate fit; each of the
clusters had a slightly different magnitude profile.

So that the model took into account this range of absolute magnitudes,
we convolved the model with this magnitude distribution, adjusting the
amplitude $A$ so that the sum of the bins in the convolution kernel was
1.0.  This way, the overall normalization of the model was not affected.

\section{Minimum $\chi^2$ fits}

Once a set of data was extracted and a set of models was defined, we fit
the models to the data by adjusting the model parameters to produce the
minimum $\chi^2$.  We considered five models: a triaxial Hernquist 
model in which the center was not fixed to the center of the Galaxy, a 
triaxial Hernquist model in which the center was forced to the center of 
the Galaxy, a triaxial power law model in which the center was allowed 
to float, a triaxial power law model in which the center was forced to the 
center of the Galaxy, and a traditional axisymmetric power law model.
The models contained between 4 and 10 free parameters, depending on the model.

For each model, we chose a starting set of parameters, excluding the
normalization $\rho_0$, and generated the expected star counts for
each bin in the star count data.  We then normalized the model so that
the sum of all of the star counts equaled the sum of all of the star
counts in the data set (including all stripes) that we were fitting.
This had the advantage of reducing by one the number of parameters that
were allowed to freely vary, since this one could be computed in a
straightforward manner.  The value of $\chi^2$ for this model
and this parameter set is then:
\[\chi^2 = \frac{\sum_{i=0}^{N_{bin}} (d_i-m_i)^2 / m_i}{N_{bin}-N_{param}-2},\]
where $N_{bin}$ is the number of bins used in all stripes, $d_i$ represents
each bin of the data, $m_i$ represents the expected number of counts
in that bin in the model, and $N_{param}$ is the number of parameters
fit in the model.  We have assumed that the statistical errors, $\sigma_i$, in the
number of stars in the bins are Poisson so that $\sigma_i^2 = m_i$.

We changed the parameters in each model by hand until we achieved the minimum
$\chi^2$ for each model.  The final parameters were checked to guarantee that
the fit is worse if any of the parameters are changed by $\pm$ the tolerance
(given in column 2 of table \ref{modelfits}) for that parameter.  If there are
10 adjustable parameters, then $3^{10}$ parameter sets were checked to ensure
a minimum.

The best fit model parameters are given in table \ref{modelfits}.
Parameters include: $R_0$, distance from 
the Sun to the Galactic center; $(p,q)$, the ratio of the scale length in
the $(Y',Z')$ directions compared to the major axis in the $X'$ direction; 
$R_{core}$, the Hernquist core radius; 
$M_f$, the peak of the absolute magnitude distribution of our turnoff 
stars; and $\rho_0$, the normalization of the Hernquist or power law model.
$(\alpha, \delta)$ are $(1,3)$
for a Hernquist profile and $(\alpha,0)$ for a power law profile.
To get from Galactic XYZ coordinates to the
principle axes of the triaxial models, we first shift the center by
$(dX,dY,dZ)$, then rotate by $\theta$ around the $Z$-axis, then rotate by 
$\phi$ around the new $Y$-axis, and then by $\xi$ around the new 
$X$-axis. 
$N_{solar}$ is the density of the selected turnoff stars in the solar
neighborhood and $\chi^2$, a measure of the goodness of the model fit.  

Numbers in bold were set to a fixed value and were not allowed to vary.  
Parameters were fixed either because the definition of the model required it 
(for example, power laws all have $\delta=0$), or because the parameter was 
correlated with a parameter that was varied.  For instance, in the full 
Hernquist model we set the scale of the model with $R_{core}$, though  we
could instead have varied $M_f$.  $R_0$ and $dX$ are redundant.

\begin{table}
\caption{\label{modelfits} Best Fit parameters for five spheroid models 
described in Section 2.  Numbers in bold were fixed; all others were varied
to minimize $\chi^2$ fit between the model and the data 
The units of $\rho_0$ are kpc$^4$ for Hernquist models, and 
kpc$^\alpha$ for power law models.}
\begin{center}
\begin{tabular}{lllllll}
\br
 & Tol. & Full & Galactocentric & Full & Galactocentric & Standard \\
       &      & Hernquist & Hernquist & Power Law & Power Law & Power Law\\
\mr
$R_0$ & 0.1 kpc & {\bf 8.0 kpc} & 8.5 kpc & {\bf 8.0 kpc} & 8.5 kpc & 10.7 kpc\\
$p$   & 0.01  & 0.73          & 0.73    & 0.74          & 0.72    & {\bf 1.0}\\
$q$   & 0.01  & 0.67          & 0.60    & 0.66          & 0.59    & 0.63\\
$\theta$ & 1$^\circ$ & 48$^\circ$ & 70$^\circ$ & 52$^\circ$ & 72$^\circ$ & --\\
$R_{core}$ & 0.5 kpc & 15.0 kpc & 14.0 kpc & --           & --      & --\\
$dX$  & 0.1 kpc & 0.1 kpc     & --      & 0.2 kpc       & --      & --\\
$dY$  & 0.1 kpc & 3.5 kpc     & --      & 3.0 kpc       & --      & --\\
$dZ$  & 0.1 kpc & 0.1 kpc     & --      & 0.0 kpc       & --      & --\\
$\phi$ & 0.5$^\circ$ & -8.0$^\circ$ & -4.5$^\circ$ & -6.5$^\circ$ & -4.0$^\circ$ & --\\
$\xi$ & 2$^\circ$ & 12$^\circ$ & 14$^\circ$ & 16$^\circ$ & 14$^\circ$ & --\\
$\alpha$ & 0.1 & {\bf 1}      & {\bf 1}  & 2.9           & 3.0     & 3.1\\
$\delta$ & -- &{\bf 3}       & {\bf 3}  & {\bf 0}       & {\bf 0} & {\bf 0}\\
$M_f$ & 0.1   & {\bf 4.2}    & {\bf 4.2} & {\bf 4.2}   & {\bf 4.2} & {\bf 4.2}\\
$\rho_0$ &    & $1.548 \times 10^8$ & $2.093 \times 10^8$ & $1.065 \times 10^6$ & $2.157 \times 10^6$ & $2.390 \times 10^6$\\
$N_{solar}$ & & 1081 kpc$^{-3}$ & 1096 kpc$^{-3}$ & 1412 kpc$^{-3}$ & 1341 kpc$^{-3}$ & 1539 kpc$^{-3}$\\
$\chi^2$ && 1.37       & 1.42    & 1.49          & 1.51    & 1.92\\
\br
\end{tabular}
\end{center}
\end{table}

\begin{figure}
\begin{minipage}{2.05in}
\includegraphics[width=2.05in]{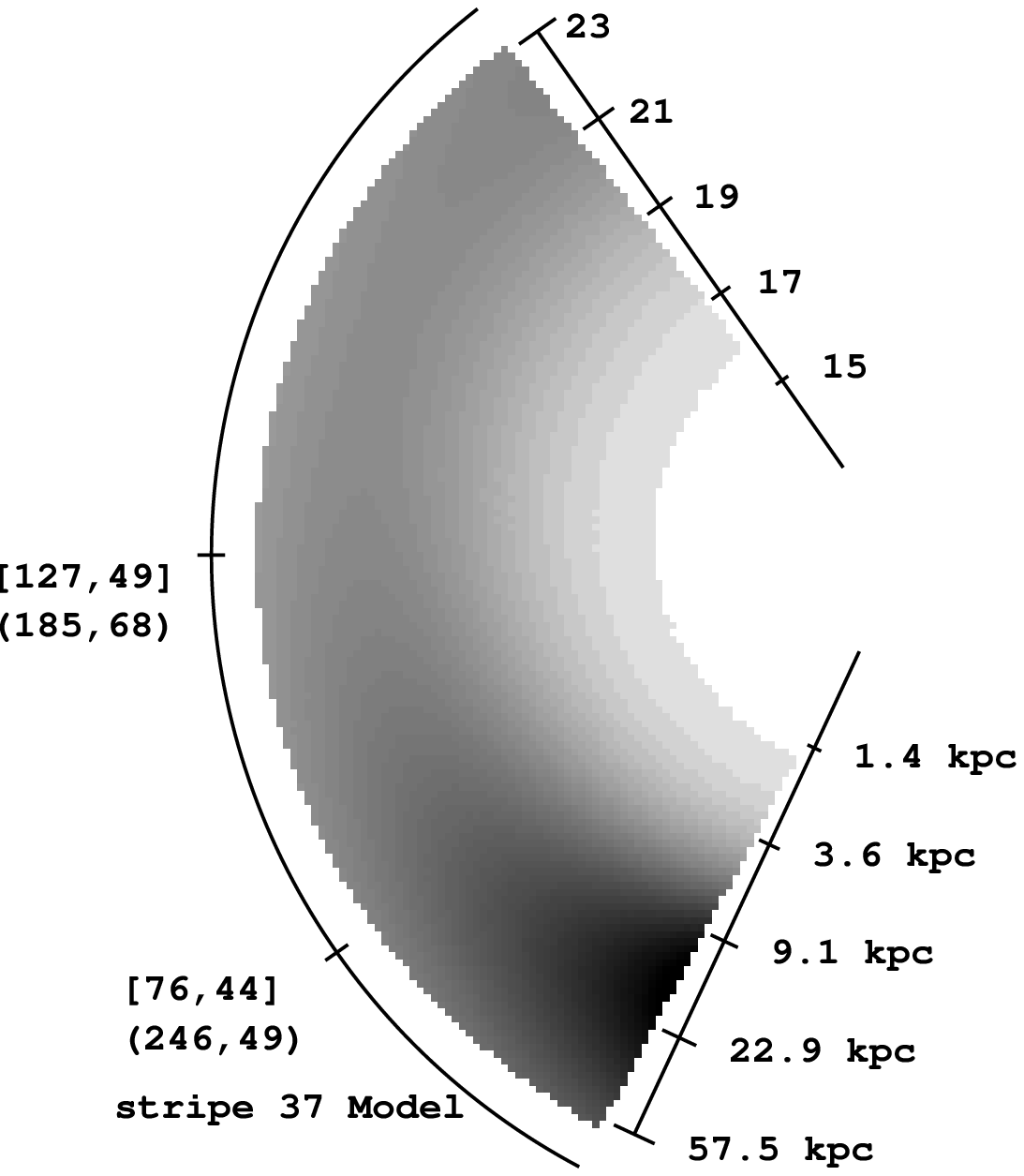}
\end{minipage}\hspace{2pc}%
\begin{minipage}{3.95in}
\includegraphics[width=3.95in]{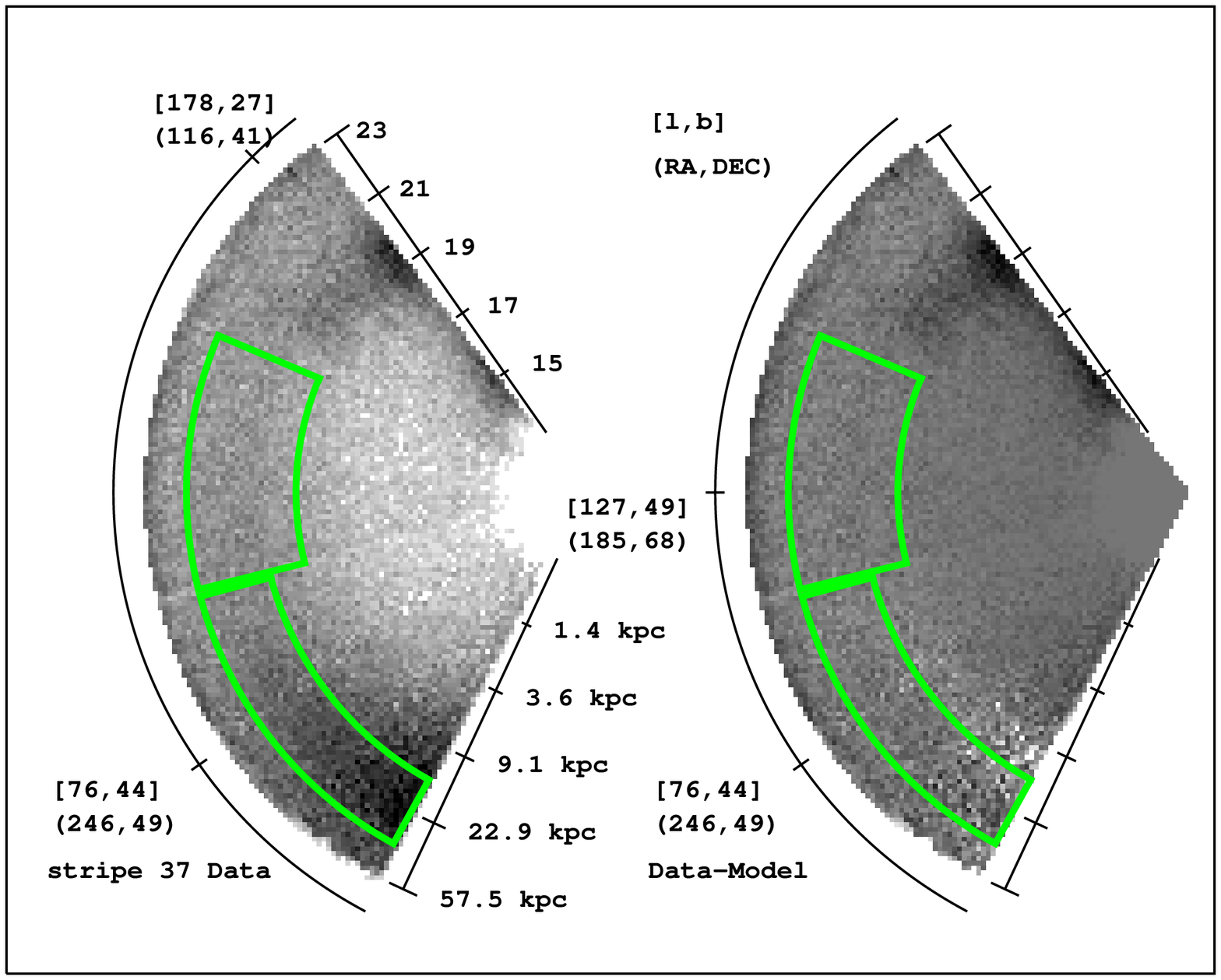}
\end{minipage}
\caption{\label{stripe37}Stripe 37 model, data, and subtraction.  The
radial coordinate is the reddening corrected $g_0$ magnitude; if we assume
the turnoff stars all have $M_{g_0} = 4.2$, then the radial coordinate
is logarithmically proportional to distance from the Sun.  The approximate
distances in kpc are given on the bottom half of the model and data plots.  The 
model and the data are shown with the same grey scale.  The two outlined 
regions on the data and the subtractions show the portions of the data on 
this stripe that were used to fit the model.  We avoided the faintest 
stars because they might not be complete, the brightest stars because they 
could be contaminated by thick disk stars, and the portion of the data at 
the top of the diagram that includes a known overdensity in the Galactic 
plane \cite{newberg2002}.  The scale for the model and data is such that 
zero counts is white; in the subtraction zero counts is grey so that 
negative deviations (white) and positive deviations (black) are both 
discernable.}
\end{figure}

\begin{figure}
\begin{center}
\includegraphics[width=3.95in]{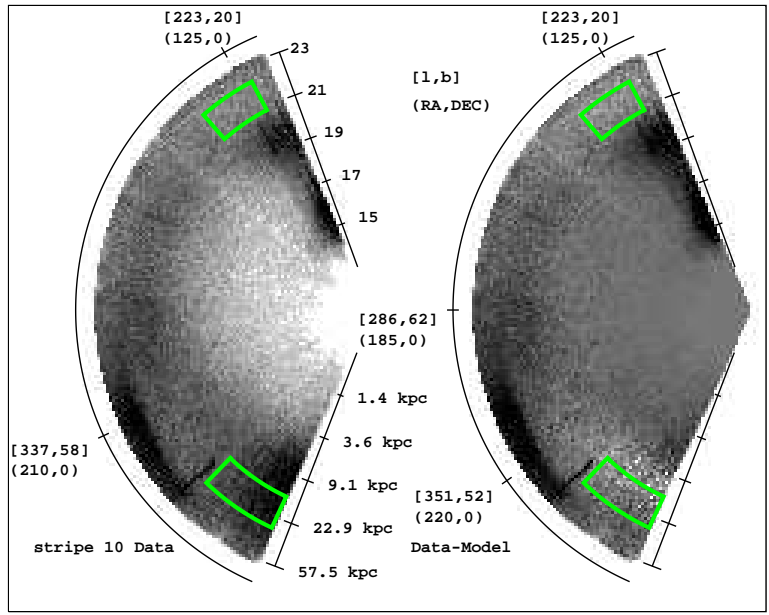}
\end{center}
\caption{\label{stripe10}Stripe 10 data and subtraction.  Very little of 
this stripe could be used for comparison due to the tidal stream in the 
plane near the top of the stripe, the Sagittarius stream on the bottom left, 
and an apparent overdensity of unknown origin on the middle left.  The 
outlined region on the top of the stripe was included because some of 
the model fits tended to oversubtract this region.}
\end{figure}

\begin{figure}
\begin{center}
\includegraphics[width=3.95in]{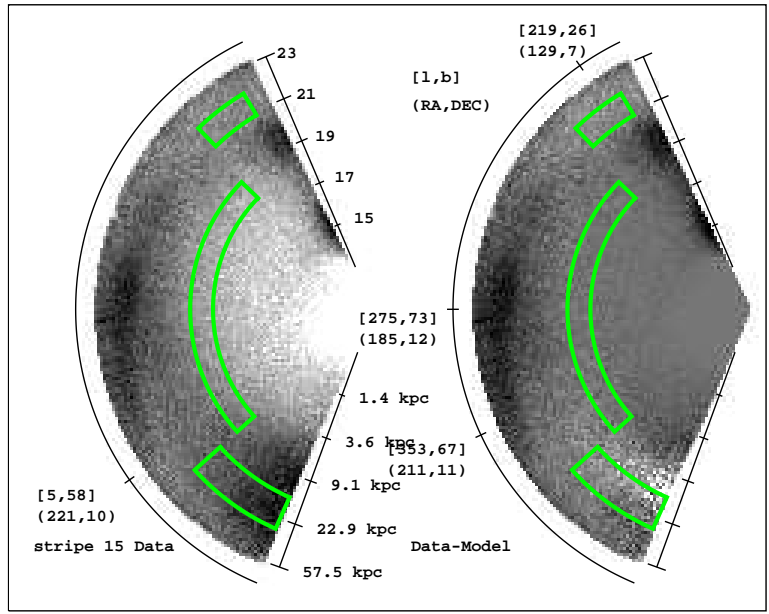}
\end{center}
\caption{\label{stripe15}Stripe 15 data and subtraction.  Prominent again in 
this figure are the Sagittarius dwarf tidal stream (left) and a tidal stream
in the Galactic plane (top).  The excess counts in the subtraction near
the bottom of the wedge, near the Galactic center, could include stars leaking
into the sample from the thick disk.}
\end{figure}

\begin{figure}
\begin{center}
\includegraphics[width=3.95in]{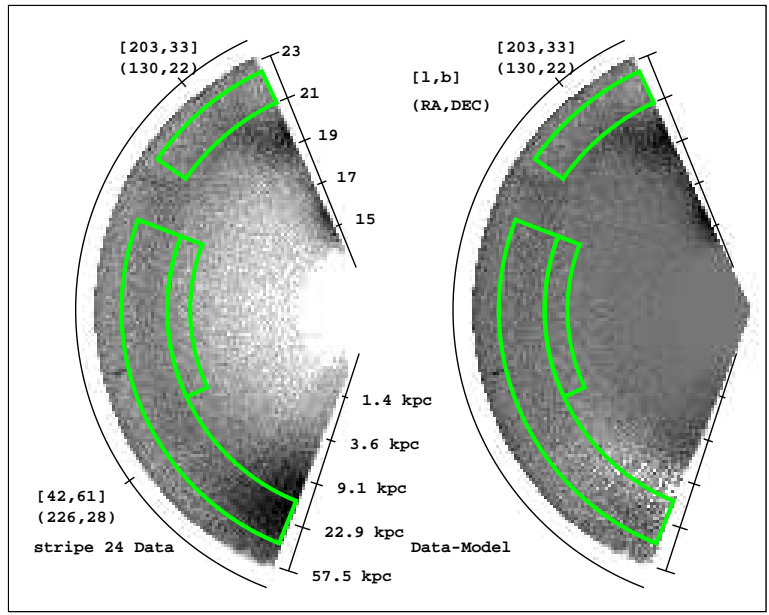}
\end{center}
\caption{\label{stripe24}Stripe 24 data and subtraction.  This stripe has
fewer large tidal streams.  The stream in the Galactic plane is still present
at the top of the diagram, and there is a small overdensity of unknown origin
about $30^\circ$ away in the counterclockwise direction.  Again, brighter stars
near the Galactic center were avoided in the fitting procedure due to
possible contamination from the thick disk.}
\end{figure}

\begin{figure}
\begin{center}
\includegraphics[width=3.95in]{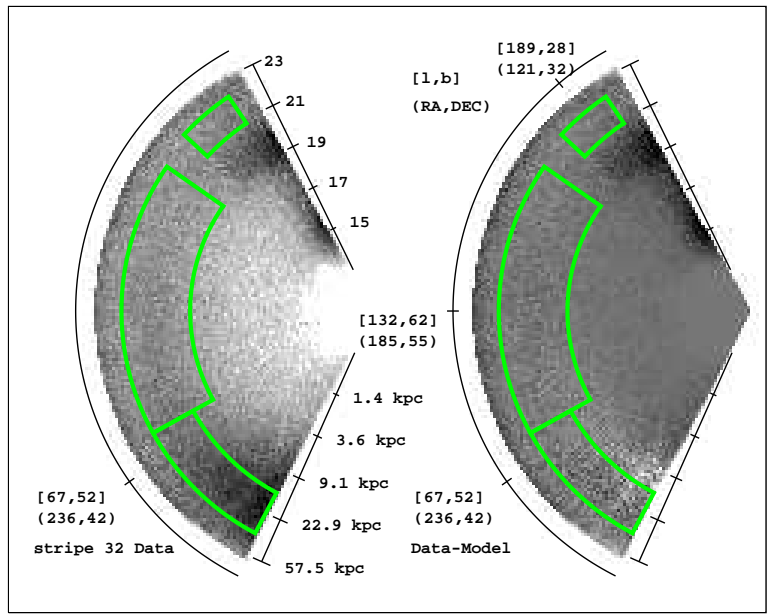}
\end{center}
\caption{\label{stripe32}Stripe 32 data and subtraction.  This is a very clean
stripe, with a nice subtraction.  The stream in the plane of the Milky Way
is present at the top of the figure.}
\end{figure}

\begin{figure}
\begin{center}
\includegraphics[width=3.95in]{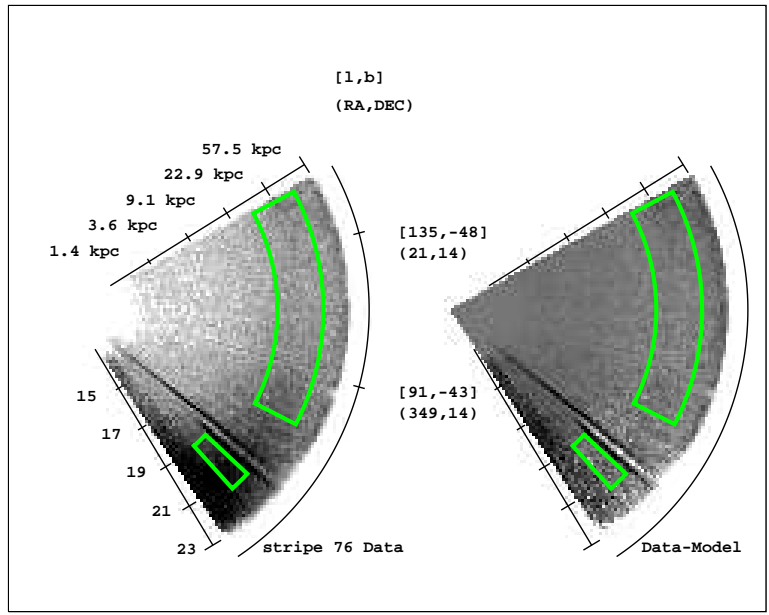}
\end{center}
\caption{\label{stripe76}Stripe 76 data and subtraction.  This is a shorter
section of data that is free from tidal debris, but contains a section of poor
data and a globular cluster (M15) that appear as dark radial features.}
\end{figure}

\begin{figure}
\begin{center}
\includegraphics[width=3.95in]{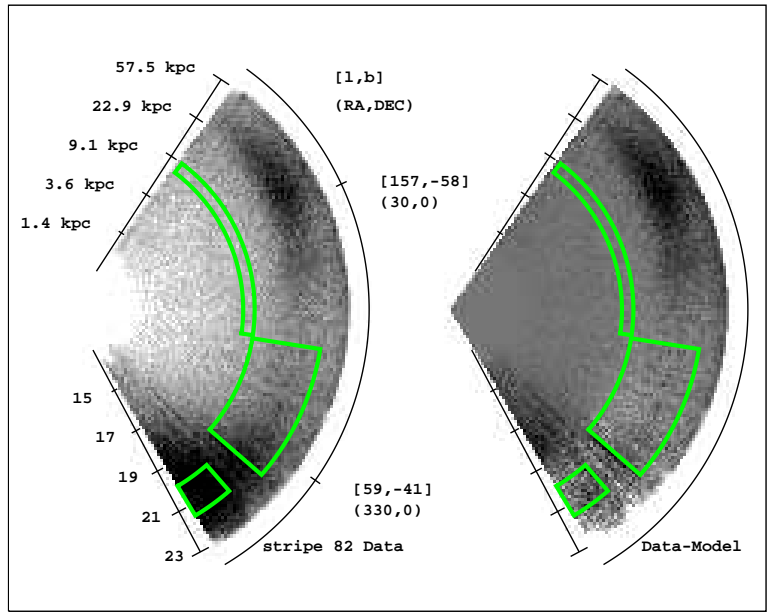}
\end{center}
\caption{\label{stripe82}Stripe 82 data and subtraction.  The trailing
tail of the Sagittarius tidal stream is prominent here.  Also, there is
a section of poor data and a globular cluster (M2) that show up as radial features
most easily seen in the subtraction.  The poor subtraction near the Galactic
center (bottom) at bright magnitudes could be due to thick disk stars
leaking into the sample.}
\end{figure}

\begin{figure}
\begin{center}
\includegraphics[width=3.95in]{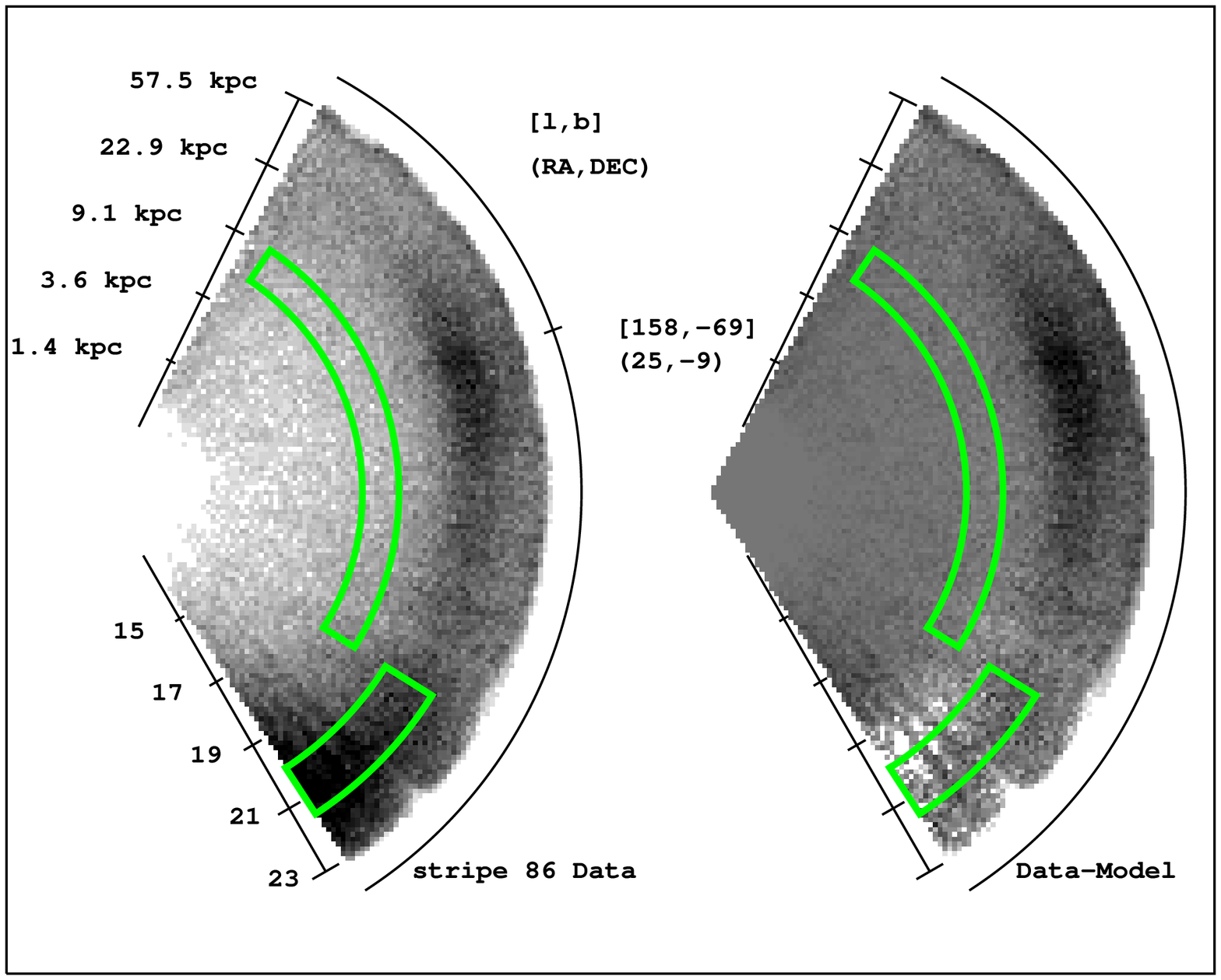}
\end{center}
\caption{\label{stripe86}Stripe 86 data and subtraction.  The Sagittarius
stream trailing tidal tail is also prominent in this stripe.  The data near
the Galactic center shows a radial pattern that is indicative of poorer
photometric data.  We experimented with leaving this stripe in and taking
it out.  In the end we decided to keep the data since the fits to this
stripe were fairly consistent with the other southern stripes, and keeping
it improved the overall appearance of the subtractions.}
\end{figure}

Figures \ref{stripe37}, \ref{stripe10}, \ref{stripe15}, \ref{stripe24},
\ref{stripe32}, \ref{stripe76}, \ref{stripe82}, and \ref{stripe86} show
the star density in the data and the residual star densities in each
stripe of data after the model has been subtracted.  Figure 
\ref{stripe37} additionally shows the star density of the model for comparison.

A crude estimate for the mass density of stars in the solar neighborhood
is calculated from SDSS measurements of the Pal 5 globular cluster.
There are 588 stars above background with $0.0 < g_0-r_0 < 0.3$
in Pal 5.  The mass of Pal 5 is 4.5 to $6.0 \times 10^3 {\rm M}_\odot$
\cite{odenkirchen2002}.  Assuming that Pal 5 is a good approximation to
the stellar population of the spheroid (a big assumption), an estimate
of the number of solar masses per sample star is $10 {\rm M}_\odot/$star.
Therefore, the local mass density of the stellar spheroid is 11,000 
M$_\odot/$kpc$^3$ for the full Hernquist model, and 15,000 M$_\odot/$kpc$^3$
for the best fit standard power law.  These numbers are 20-25 times lower
than the average density from {\it Allen's Astrophysical Quantities}
\cite{allen} of $2.6 \times 10^5 {\rm M}_\odot/{\rm kpc}^3$.  Since Pal
5 has a bluer turnoff than the spheroid, it is likely that using this
cluster to estimate the mass of the spheroid component leads to an
underestimate, though it seems unlikely that it could account for the
whole difference.  Future work will provide a better estimate of the 
spheroid normalization.

\section{Conclusions} 

The standard stellar density profile used to fit the Galactic spheroid,
$\rho = \rho_0 / r^\alpha,$ is an extremely poor fit to the density of
halo turnoff stars when a large area of the sky is sampled, even using  
the unusually large best fit distance between the Sun and the Galactic 
center.  The flattening of $q = 0.63$ and the power law slope of 
$\alpha = 3.1$ are within the range of previously measured values.  

Triaxial models are best fit with a major axis $50-70^\circ$ from the
line of sight from the Sun to the center of the Galaxy and tilted 
$4^\circ-6^\circ$
from the Galactic plane.  The minor axis, $\approx 13^\circ$ from the 
$Z$-direction, is about $65\%$ of the scale length of the major axis.  
The intermediate axis is about $75\%$ of the scale length of the major axis, 
and tilted at about $12^\circ$ from the Galactic plane.  These models
are clearly favored over the axisymmetric standard model.

The Hernquist profile is a better fit to the data than a power law
profile, but give roughly similar values for the positional and flattening
parameters.  All of the models suggest a lower spheroid stellar density 
near the Sun compared with previous estimates.

If the center of the spheroid profile is allowed to vary, then it moves
a surprising 3.0 - 3.5 kpc from the nominal center of the disk in the
direction of the Sun's motion.  Since this is perpendicular to the line of
sight from the Sun to the Galactic center, it is not possible to reduce this
offset by simply changing the scale lengths or assumed absolute magnitudes of the 
sample stars.  It is the stars in the southern stripes near the Galactic
center that tend to pull the spheroid density off center.  We are still
investigating whether there could be unidentified substructure or data
quality issues at the root of this surprising result.  Clearly, more data
in the southern hemisphere would be useful in understanding the large-scale
distribution of stars in the halo.

A large fraction of the Galactic halo contains known substructure that
must be avoided to fit the smooth component.  It is unknown whether the
results provided here are contaminated by unidentified substructure.

\ack
H. J. N. acknowledges funding from Research Corporation and the National 
Science Foundation (AST 03-07571).  Ken Freeman and Markus Samland provided
stimulating discussion of the of the stellar content of the Milky Way.

Funding for the creation and distribution of the SDSS Archive has been provided by the Alfred P. Sloan Foundation, the Participating Institutions, the National Aeronautics and Space Administration, the National Science Foundation, the U.S. Department of Energy, the Japanese Monbukagakusho, and the Max Planck Society. The SDSS Web site is http://www.sdss.org/.

The SDSS is managed by the Astrophysical Research Consortium (ARC) for the Participating Institutions. The Participating Institutions are The University of Chicago, Fermilab, the Institute for Advanced Study, the Japan Participation Group, The Johns Hopkins University, the Korean Scientist Group, Los Alamos National Laboratory, the Max-Planck-Institute for Astronomy (MPIA), the Max-Planck-Institute for Astrophysics (MPA), New Mexico State University, University of Pittsburgh, University of Portsmouth, Princeton University, the United States Naval Observatory, and the University of Washington.

\smallskip


\begin{thebibliography}{9}
\bibitem{sfd} Schlegel D J, Finkbeiner D P and Davis M 1998 {\it Astrop. J.} {\bf 500} 525
\bibitem{dr1} Abazajian K et al. 2003 {\it Astron. J.} {\bf 126} 2081
\bibitem{dr2} Abazajian K et al. 2004 {\it Astron. J.} {\bf 128} 502
\bibitem{dr3} Abazajian K et al. 2005 {\it Astron. J.} {\bf 129} 1755
\bibitem{fukugita} Fukugita M, Ichikawa T, Gunn J E, Doi M, Shimasaku K and Schneider D P 1996, {\it Astron. J.} {\bf 111} 1748
\bibitem{gunn} Gunn J E et al. 1998 {\it Astron. J.} {\bf 116} 3040
\bibitem{hogg} Hogg D W, Finkbeiner D P, Schlegel D J and Gunn J E 2001, {\it Astron. J.} {\bf 122} 2129
\bibitem{stoughton} Stoughton C et al. 2002 {\it Astron. J.} {\bf 123} 485
\bibitem{pier} Pier J R, Munn J A, Hindsley R B, Hennessy G S, Kent S M, Lupton R H and Ivezic Z 2003 {\it Astron. J.} {\bf 125} 1559
\bibitem{smith} Smith J A et al. 2002 {\it Astron. J.} {\bf 123} 2121
\bibitem{york} York D G et al. 2000 {\it Astron. J.} {\bf 120} 1579
\bibitem{astrom} Newberg H J and Yanny B 2005 {\it Astrometry in the Age of the Next Generation
of Large Telescopes (18-20 October 2004, Flagstaff, AZ)} vol 338 ed A  K B Monet and K Seidelmann (San Francisco: PASP Conf. Ser.) in press ({\it Preprint} astro-ph/0502386)
\bibitem{besancon} Robin A C, Reyle S D and Picaud S 2003 {\it Astron. Astrophys.} {\bf 409} 523
\bibitem{newberg2002} Newberg, H J et al. 2002 {\it Astroph. J.} {\bf 569} 245
\bibitem{odenkirchen2002} Odenkirchen M, Grebel E K, Dehnen W, Rix H-W and 
Cudworth K M 2002 {\it Astron. J.} {\bf 124} 1497
\bibitem{allen} Cox A 1999 {\it Allen's Astrophysical Quantities} fourth
edition (New York: Springer-Verlag) chapter 23 p 571
\end{thebibliography}
\end{document}